\documentclass{llncs}
\usepackage{amsmath}
\usepackage{amssymb}
\usepackage{enumerate}
\usepackage{amsfonts}
\usepackage{graphicx}
\usepackage{eufrak}

\usepackage{xspace}
\usepackage{mathrsfs}
\usepackage{verbatim}
\usepackage{setspace}
\usepackage{tabularx}
\usepackage{color}
\usepackage{url}

\newcounter{instr}

\newcommand{\best}[1]{\underline{\textbf{#1}}}
\newcommand{\good}[1]{\textbf{#1}}

\newcommand{\sht}{|}

\begin{document}

\title{Flexible and Efficient Algorithms\\for Abelian Matching in Strings\thanks{This is a short preliminary version of a full paper submitted to an international journal. Most examples, details, lemmas and theorems have been omitted.}}

 \author{Simone Faro$^{\dagger}$ and Arianna Pavone$^{\ddagger}$}
 
 \institute{
 $^{\dagger}$Universit\`a di Catania, Dipartimento di Matematica e Informatica\\Viale Andrea Doria 6, I-95125 Catania, Italy\\\email{faro@dmi.unict.it}\\[0.2cm]
 $^{\ddagger}$Universit\`a di Messina, Dipartimento di Scienze Cognitive\\Via Concezione 6, I-98121 Messina, Italy\\\email{apavone@unime.it} 
 }

\maketitle


\begin{abstract}
The \emph{abelian pattern matching problem} consists in finding all substrings of a text which are permutations of a given pattern.  This
problem finds application in many areas and can be solved in linear time by a na\"{i}ve sliding window approach.  
In this short communication we present a new class of algorithms based on a new efficient fingerprint computation approach, called \emph{Heap-Counting}, which turns out to be fast, flexible and easy to be implemented. It can be proved that our solutions have a linear worst case time complexity and, in addition, we present an extensive experimental evaluation which shows that our newly presented algorithms are among the most efficient and flexible solutions in practice for the abelian matching problem in strings.
\end{abstract}

\section{Introduction}\label{sec:Introduction}

Given a pattern $x$ and a text $y$, the \emph{abelian pattern matching} problem~\cite{Ejaz10} (also known as \emph{jumbled matching}~\cite{BCFL12b,BCFL12,Ghu16} or \emph{permutation matching} problem) is a well known special case of the approximate string matching problem and consists in finding all substrings of $y$, whose characters have the same multiplicities as in $x$, so that they could be converted into the input pattern just by permuting their characters.


This problem naturally finds applications in many areas, such as string alignment~\cite{Ben03}, SNP discovery~\cite{Bocker07}, and also in the interpretation of mass spectrometry data~\cite{Bocker04}. We refer to the recent paper by  Ghuman and  Tarhio \cite{Ghu16} for a detailed and broad list of applications of the abelian pattern matching problem.

In addition abelian pattern matching is also wisely used in the field of text processing and in computational biology, where algorithms for such problem are used as a filtering technique~\cite{BYN02}, usually referred to as {\em counting filter}, to speed up complex combinatorial searching problems.  The basic idea is that in many approximation problems a substring of the text which is an occurrence of a given pattern, under a specific distance function, is also a permutation of it.
For instance, the counting filter technique has been used solutions to the approximate string matching problem allowing for mismatches~\cite{GL89}, differences~\cite{JTU96}, inversions~\cite{CCF11} and translocations~\cite{GFG11}.

\smallskip

In this paper we are interested in the \emph{online} version of the problem which assumes that the input pattern and text are given together for a single instant query, so that no preprocessing of the text is possible. Although its worst-case time complexity is well known to be
$O(n)$, in the last few years much work has been made in order to speed up the performances of abelian matching algorithms in practice, and some very efficient algorithms have been presented, tuned for specific settings of the problem \cite{CGT15,Ghu16}.

We present two algorithms based on a new efficient fingerprint computation approach, called \emph{Heap-Counting}, which turns out to be fast, flexible and ease to be implemented. The first algorithm is designed using a prefix based approach, while the second one uses a suffix based approach. It can be proved that both of them have a linear worst case time complexity.

In addition we present also two fast variants of the above algorithms, obtained by relaxing some algorithmic constraints, which, despite their quadratic worst case time complexity, turn out to be faster in some specific practical cases.

From our experimental results it turns out that our newly presented algorithms are among the most efficient and flexible solutions for the abelian matching problem in strings.

\smallskip

The paper is organized as follows.
After introducing in Section~\ref{sec:notations} the relevant notations, we present in Section~\ref{sec:new} two new solutions of the online abelian pattern matching problem in strings, based on the Heap-Counting approach. Then, in Section~\ref{results}, we present a detailed experimental evaluation of the new presented algorithms, comparing them against the most effective solutions known in literature.

\section{Basic Notions}\label{sec:notations}
Before entering into details we recall some basic notions and introduce some useful notations.

We represent a string $x$ of length $|x| = m > 0$ as a finite array $x[0 \,..\, m-1]$ of characters from a finite alphabet $\Sigma$ of size $\sigma$.  Thus, $x[i]$ will denote the $(i+1)$-st character of $x$, for $0\leq i < m$, whereas $x[i\, ..  \, j]$ will denote the substring of $x$ contained between the $(i + 1)$-st and the $(j + 1)$-st characters of $x$.

Given a string $x$ of length $m$, the occurrence function of $x$, $\rho_x: \Sigma \rightarrow \{0, ..., m\}$, associates each character of the alphabet with its number of occurrences in $x$. Formally, for each $c \in \Sigma$, we have:
$$
	\rho_x(c) = \big| \big\{ i : x[i]=c \big\}\big|.
$$

The \emph{Parikh vector}~\cite{AALS03,Sal03} of $x$ (denoted by $pv_{x}$ and also known as \emph{compomer} \cite{Bocker04}, \emph{permutation pattern} \cite{ELP04}, and \emph{abelian pattern} \cite{Ejaz10}) is the vector of the multiplicities of the characters
in $x$.  More precisely, for each $c \in \Sigma$, we have 
\[
pv_{x}[c] = \big|\big\{i : 0 \leq i < m \text{ and } x[i]=c\big\}\big|\,.
\]
In the following, the Parikh vector of the substring $x[i \,..\, i+h-1]$ of $x$, of length $h$ and starting at position $i$, will be denoted by $pv_{x(i,h)}$.

Fig. \ref{code:auxiliary} shows the pseudocode of procedure \textsc{Compute-Parikh-Vector} for computing the Parikh vector $pv_x$ of a string $x$ of length $m$. It needs an initialization of the vector (line 1) which takes $O(\sigma)$ time, and an inspection of all characters of $x$ (line 2) which takes $O(m)$ time. Thus the Parikh vector can be computed in $O(m+\sigma)$ time.

In terms of Parikh vectors, the abelian pattern matching problem can be formally expressed as the problem of finding the set
$\Gamma_{x,y}$ of positions in $y$, defined as
$$
	\Gamma_{x,y} = \big\{ s : \ 0\leq s \leq n-m \textrm{ and }
	pv_{y(s,m)} = pv_{x} \big\}.
$$

Thus a first brute-force solution for the abelian pattern matching problem consists in checking for all possible text positions $s$, with $0\leq s < n-m$, if the relation $pv_{y(s,m)} = pv_{x}$ holds. Fig. \ref{code:auxiliary} shows the pseudocode of such brute-force algorithm and its auxiliary procedures. Specifically it makes use of procedure \textsc{Verify} which checks if the substring of the text, with length $m$ and  starting at position $s$, has the same Parikh vector as $x$. procedure \textsc{Verify} takes $O(\sigma+m)$ time in the worst case, thus the overall worst case time complexity of the brute-force algorithm is $O(n(\sigma+m))$.

\begin{figure}[!t]
\begin{center}
\begin{tabular}{ll}
	\multicolumn{2}{l}{\textsc{Compute-Parikh-Vector}($x$, $m$)}\\ 
	1. & \quad \textsf{for each $c \in \Sigma$ do $pv_x[c] \leftarrow 0$ }\\
	2. & \quad \textsf{for $i \leftarrow 1$ to $m$ do $pv_x[x[i]] \leftarrow pv_x[x[i]]+1$ }\\
	3. & \quad \textsf{return $pv_x$ }\\
	&\\
	\multicolumn{2}{l}{\textsc{Verify}($pv_x$, $m$, $y$, $s$)}\\ 
	1. & \quad \textsf{for each $c \in \Sigma$ do $pv_{y(s,m)}[c] \leftarrow 0$ }\\
	2. & \quad \textsf{for $i \leftarrow 1$ to $m$ do}\\
	3. & \quad \qquad \textsf{$pv_{y(s,m)}[y[s+i]] \leftarrow pv_{y(s,m)}[y[s+i]]+1$ }\\
	4. & \quad \qquad \textsf{if $pv_{y(s,m)}[y[s+i]]>pv_x[y[s+i]]$ return False}\\
	5. & \quad \textsf{return True }\\
	&\\
	\multicolumn{2}{l}{\textsc{Brute-Force-Abelian-Matching}($x$, $m$, $y$, $n$)}\\ 
	1. & \quad \textsf{$pv_x \leftarrow$ \textsc{Compute-Parikh-Vector($x$,$m$)}}\\
	2. & \quad \textsf{for $s \leftarrow 0$ to $n-m$ do}\\
	3. & \quad \qquad \textsf{if \textsc{Verify}($pv_x, m , y, s$) then}\\
	4. & \quad \qquad \qquad \textsf{Output($s$)}\\
\end{tabular}
\end{center}
\caption{\label{code:auxiliary} (1) Procedure \textsc{Compute-Parikh-Vector}, for computing the Parikh vector of a string $x$ of length $m$; (2) Procedure \textsc{Verify}, for testing is the substring of the text beginning at position $s$ is an abelian occurrence of the pattern $x$ of length $m$; (3) Procedure \textsc{Brute-Force-Abelian-Matching}, a naive algorithm for the abelian matching problem, working in $O(n(\sigma+m))$ worst case time complexity.}
\end{figure}



\section{The Heap-Counting Approach}\label{sec:new}
Let $x$ and $y$ be strings of length $m$ and $n$, respectively, over a common alphabet $\Sigma$ 
of size $\sigma$.  
Previous solutions for the abelian pattern matching problem maintain in constant time the symmetric difference, $e$, of the multisets of the characters occurring in the current text window and of those occurring in the pattern, respectively.  Thus, when $e= 0$, a match is reported.  Alternatively, they use a packed representation of the Parikh vector, where some kind of overflow sentinel is implemented in order to take track that the frequency of a given character has exceeded its corresponding value in the Parikh vector of the pattern.
The aim is to perform the inizialization of the Parikh vector in constant time and to perform vector updates in a very fast way.

Instead of using a structured representation of the Parikh vector of a string, fitting in a single computer word, our approach tries to map the multisets of our interest into natural numbers, using a heap-mapping function $h$ that allows for very fast updates.

Specifically we suppose to have a function $h:\Sigma \rightarrow \mathbf{N}$ (the \emph{heap-function}), that maps each character $c$ of the alphabet $\Sigma$ to a natural number, $h(c)$ indeed. Then, we assume that the multiset of a given string $w\in \Sigma^*$, of length $m$, can be univocally associated to a natural number, $h(w)$, using the following relation:

\begin{equation}
	h(w) = \sum_{i=0}^{m-1} h(w[i])
\end{equation}
The value $h(w)$ is called the \emph{heap-value} of the string $w$. In this context a abelian match is found at position $s$ of the text when the heap-value associated to the window starting at position $s$  is equal to the heap-value of the pattern. This approach, when applicable, leads to two main advantages: the multisets of the characters occurring in string can be represented by a single numeric value, fitting in a single computer word; modifications and updates of such multisets can be done by means of simple integer additions.

Our heap-counting approach is based on the following elementary fact.

\begin{lemma}\label{lem:permutations}
Let $\Sigma = \{c_0, c_1, \ldots, c_{\sigma-1}\}$ be an alphabet of
size $|\Sigma|=\sigma$, let $m>1$ be an integer, and let
$h:\Sigma \rightarrow \mathbb{N}$ be the mapping $h(c_i) =
m^{i}$, for $i=0, \ldots, \sigma-1$. 
Then for any two distinct $k$-multicombinations (i.e.,
$k$-combinations with repetitions) $\varphi_1$ and $\varphi_2$ from
the set $\Sigma$, with $1 \leq k\leq m$, we have
\begin{equation}
    \label{eq_star}
    \sum_{c\in \varphi_1} h(c) \neq \sum_{c\in \varphi_2} h(c)\,.
\end{equation}\qed
\end{lemma}

\begin{figure}[!t]
\begin{center}
\begin{tabular}{ll}
	\multicolumn{2}{l}{\textsc{Compute-Heap-Mapping}($x$, $m$)}\\ 
	1. & \quad \textsf{for each $c \in \Sigma$ do $h[c] \leftarrow 0$ }\\
	2. & \quad \textsf{$j \leftarrow 1$}\\
	3. & \quad \textsf{for $i \leftarrow 1$ to $m$ do}\\
	4. & \quad \qquad \textsf{if $h(x[i]) = 0$ then}\\
	5. & \quad \qquad \qquad \textsf{$h(x[i]) \leftarrow j$ }\\
	6. & \quad \qquad \qquad \textsf{$j \leftarrow j\times m$ }\\
	7. & \quad \textsf{return $h$ }\\
	&\\
	\multicolumn{2}{l}{\textsc{Heap-Counting-Abelian-Matching}($x$, $m$, $y$, $n$)}\\ 
	1. & \quad \textsf{$h \leftarrow $\textsc{Compute-Heap-Mapping}($x$, $m$)}\\
	2. & \quad \textsf{$\delta \leftarrow \gamma \leftarrow 0$}\\
	3. & \quad \textsf{for $i \leftarrow 0$ to $m-1$ do }\\
	4. & \quad \qquad \textsf{$\delta \leftarrow \delta + h(x[i])$ }\\
	5. & \quad \qquad \textsf{$\gamma_0 \leftarrow \gamma_0 + h(y[i])$ }\\
	6. & \quad \textsf{if $\gamma_0 = \delta$ then \textsc{Output}(0) }\\
	7. & \quad \textsf{for $s \leftarrow 1$ to $n-m$ do }\\
	8. & \quad \qquad \textsf{$\gamma_s \leftarrow \gamma_{s-1} + h(y[s+m-1])-h(y[s-1])$ }\\
	9. & \quad \qquad \textsf{if $\gamma_s = \delta$ \textsc{Output}($s$) }\\
\end{tabular}
\end{center}
\caption{\label{code:searching1} The pseudocode of the \textsc{Heap-Counting-Abelian-Matching} for the online exact abelian matching problem, implemented using a prefix-based approach.}
\end{figure}

Let $\diamond $ be a character such that $\diamond \notin \Sigma$ and let $\Sigma_x \subseteq \Sigma $ be the set of all (and only) the characters occurring in $x$. We indicate with $\sigma_x$ the size of the alphabet $\Sigma_x$. Plainly we have $\sigma_x\leq \min\{\sigma,m\}$, thus we can think to this transformation as a kind of alphabet reduction. 

We define the \emph{reduced text} $\bar{y}$, over $\Sigma_x$, as a version of the text $y$ where each character $y[i]$, not included in $\Sigma_x$, is replaced with the special character $\diamond  \notin \Sigma$.
Since, in general, $\sigma_x < \sigma$ (especially in the case of short patterns), to process the reduced version of the text, instead of its original version, allows the heap function to be computed on a smaller domain, reducing therefore the size of the heap-values associated to any given string.

The following Lemma \ref{lem:reduced} proves that such alphabet reduction does not influence the the output of any abelian pattern matching algorithm (the proof is omitted).

\begin{lemma}\label{lem:reduced}
Let $x$ be a string of length $m$ and $y$ be a string of length $n$, both over the alphabet $\Sigma$ of size $\sigma$. Moreover let $\Sigma_x$ be the set of characters occurring in $x$ and let $\bar{y}$ be the reduced version of $y$ over $\Sigma_x$. Then the occurrences of $x$ in $y$ correspond to the occurrences of $x$ in $\bar{y}$. Formally
$$
	\Gamma_{x,y} = \Gamma_{x,\bar{y}}
$$  \qed
\end{lemma}

\subsection{A Prefix-Based Algorithm}\label{new:prefix}
Inspired by Lemma \ref{lem:permutations} and by Lemma \ref{lem:reduced}, the new algorithm precomputes the set $\Sigma_x$ and the function $h:\Sigma_x \rightarrow \mathbb{N}$, defined as $h(c_i) = m^i$, for $i=0,\ldots, \sigma_x-1$, where $m$ is the length of the pattern and $\sigma_x$ is the size of $\Sigma_x$. 

Fig. \ref{code:searching1} shows the \textsc{Heap-Abelian-Matching} algorithm and its  the auxiliary procedure. 

During the preprocessing phase (lines 1-7) the algorithm precomputes the heap-mapping function $h$ (line 1) by means of procedure  \textsc{Compute-Heap-Mapping}. Such procedure computes the mapping table over the alphabet $\Sigma_x \cup \{\diamond\}$ associating the value $1$ with all characters not occurring in $x$, i.e. we set $h(\diamond)=1$.

The heap values $\delta = h(x)$ and $\gamma_0 = h(y[0..m-1])$ are then precomputed in lines 2-5, Likewise, during the searching phase (lines 8-10), the heap value $\gamma_{i} = h(y[s\,..\,s+m-1])$ is computed for each window $y[s\,..\,s+m-1]$ of the text $t$, with $0 < s \leq n-m$.
Specifically, starting from the heap value $\gamma_{s-1}$, the algorithm computes the heap value $\gamma_s$ by using the relation  $\gamma_s = \gamma_{s-1} - h(y[s-1]) + h(y[s+m-1])$ (line 8). Of course, in practical implementations of the algorithm it is possible to maintain a single value $\gamma$, corresponding to the heap value of the current window of the text.
 
The set $\Gamma_{x,y}$ of all occurrences in the text is then 
$$ 
\Gamma_{x,y} = \displaystyle \big\{ s\ \sht\ 0\leq s \leq n-m \textrm{ and } \gamma_{s} = \delta \big\}
$$

We can prove that the algorithm \textsc{Forward-Heap-Abelian-Matching} computes all abelian occurrences of $x$ in $y$ with $O(m+\sigma)$ -time and -space complexity in the worst case (the proof is omitted).

From a practical point of view it is understood that for an architecture, say, at
64 bits, all operations will take place modulo $2^{64}$. Thus, when $m^{\sigma_x+1}$ exceeds $2^{64}$ we could have some collisions in the set of the heap values and an additional verification procedure should be run every time an occurrence is found. 
However it has been observed experimentally that, also in this specific cases, the collision problem for the
heap function $h$ is negligible.


\subsection{A Suffix-Based Algorithm}\label{new:suffix}
In this section we extend the idea introduced in the previous section and present a backward version of the prefix-based algorithm described above which turns out to be more efficient in the case of long patterns or large alphabets. It shows a sub-linear behaviour in practice, while maintains the same worst case time complexity.
Figure \ref{code:searching2} shows the new algorithm, called \textsc{Backward-Heap-Abelian-Matching}, and its  the auxiliary procedure. 

\begin{figure}[!t]
\begin{center}
\begin{tabular}{ll}
	\multicolumn{2}{l}{\textsc{Compute-Membership-Map}($x$, $m$)}\\ 
	1. & \quad \textsf{for each $c \in \Sigma$ do $b(c) \leftarrow$ False}\\
	2. & \quad \textsf{for $i \leftarrow 1$ to $m$ do $b(x[i]) \leftarrow$ True}\\
	3. & \quad \textsf{return $b$ }\\
	& \\
	\multicolumn{2}{l}{\textsc{Backward-Heap-Counting-Abelian-Matching}($x$, $m$, $y$, $n$)}\\ 
	1. & \quad \textsf{$h \leftarrow $\textsc{Compute-Heap-Mapping}($x$, $m$)}\\
	2. & \quad \textsf{$b \leftarrow $\textsc{Compute-Membership-Mapping}($x$, $m$)}\\
	3. & \quad \textsf{$\delta \leftarrow 0$}\\
	4. & \quad \textsf{for $i \leftarrow 1$ to $m$ do $\delta \leftarrow \delta + h(x[i])$ }\\
	5. & \quad \textsf{$y \leftarrow y.x$ }\\
	6. & \quad \textsf{$s \leftarrow 0$}\\
	7. & \quad \textsf{while ( True ) do}\\
	8. & \quad \qquad \textsf{$\gamma \leftarrow -\delta$}\\
	9. & \quad \qquad \textsf{$j \leftarrow m-1$}\\
	10. & \quad \qquad \textsf{while ($j\geq 0$) do}\\
	11. & \quad \qquad \qquad \textsf{if ($b(y[s+j])$) then}\\
	12. & \quad \qquad \qquad \qquad \textsf{$\gamma \leftarrow \gamma + h(y[s+j])$}\\
	13. & \quad \qquad \qquad \qquad \textsf{$j \leftarrow j - 1$}\\
	14. & \quad \qquad \qquad \textsf{else}\\
	15. & \quad \qquad \qquad \qquad \textsf{$\gamma \leftarrow -\delta$}\\
	16. & \quad \qquad \qquad \qquad \textsf{$s \leftarrow s + j + 1$}\\
	17. & \quad \qquad \qquad \qquad \textsf{$j \leftarrow m-1$}\\
	18. & \quad \qquad \textsf{do}\\
	19. & \quad \qquad \qquad \textsf{if ($\gamma = 0$) then}\\
	20. & \quad \qquad \qquad \qquad \textsf{if ($s\leq n-m$) then \textsc{Output}($s$)}\\
	21. & \quad \qquad \qquad \qquad \textsf{else return}\\
	22. & \quad \qquad \qquad \textsf{if ($b(y[s+m])=$False) then break}\\
	23. & \quad \qquad \qquad \textsf{$\gamma \leftarrow \gamma -h(y[s]) + h(y[s+m])$}\\
	24. & \quad \qquad \qquad \textsf{$s \leftarrow s+1$}\\
	25. & \quad \qquad \textsf{while ( True )}\\
	26. & \quad \qquad \textsf{$s \leftarrow s+m+1$}\\
\end{tabular}
\end{center}
\caption{\label{code:searching2} The pseudocode of the \textsc{Backward-Heap-Counting-Abelian-Matching} for the online exact abelian matching problem, implemented using a suffix-based approach.}
\end{figure}

During the preprocessing phase (lines 1-6) the algorithm precomputes the heap-mapping function $h$ (line 1) and the membership function $b$ (line 2). We use procedure \textsc{Compute-Heap-Mapping}, and procedure \textsc{Compute-Membership-Mapping}, respectively.

The heap value $\delta = h(x)$ of the pattern $x$ is then precomputed in lines 3-4. A copy of the pattern is then concatenated at the end of the pattern (line 5), as a sentinel, in order to avoid the window of the text to shift over the last position of the text.

The main cycle of the searching phase (line 7) is executed until the value of $s$ becomes greater than $n-m$ (line 20). An iteration of the main cycle is divided into two additional cycles. The first cycle of line 10 performs a backward scanning of the current window of the text and stops when the whole window has been scanned or a character not occurring in $\Sigma_x$ is encountered.
The second cycle of line 18, starting from the heap value of the current window of the text, computes at each iteration the heap value of the next window in constant space using a forward scan. The second cycle stops when a character not occurring in $\Sigma_x$ is encountered.

It is possible to prove that the algorithm \textsc{Backward-Heap-Counting-Abelian-Matching} computes all abelian occurrences of $x$ in $y$ with $O(\sigma+m+n)$-time and $O(\sigma+m)$-space complexity in the worst case (the proof is omitted).

\subsection{Relaxed Filtering Variants}\label{new:filter}
A simpler implementation of the above presented algorithms can be obtained by relaxing the heap-counting approach presented at the beginning of this section, in order to speed-up the computation of the heap values of a string and, as a consequence, to spud-up the searching phase of the algorithm.

Specifically we propose to use the natural predisposition of the characters of an alphabet to be treated as integer numbers.
For instance, in many practical applications, input strings can be handled as sequences of ASCII characters. In such applications, characters can just be seen as the $8$-bit integers corresponding to their ASCII code. 

In this context, if we indicate with \textsc{ascii}($c$), the ASCII code of a character $c\in\Sigma$, we can set $h(c) = \textsc{ascii}(c)$. Thus the heap value of a string can be simply computed as the sum of the ASCII codes of its characters.

As a consequence the resulting algorithms works as filtering algorithm. Indeed, when an occurrence is found we are not sure that the substring of the text which perform a match is a real permutation of the pattern. This implies that an additional verification phase is run for each candidate occurrences.

Plainly the resulting algorithms have an $O(\sigma + nm)$ worst case time complexity, since a verification procedure could be run for each position of the text.

\section{Experimental Results}\label{results}
We report in this section the results of an extensive experimentation of the newly presented algorithms against the most efficient solutions known in literature for the online abelian pattern matching problem.
In particular we have compared 11 algorithms divided in three groups: prefix-based, suffix-based and SIMD based algorithms. 
Specifically we compared the following 5 prefix based algorithms:
\begin{enumerate}
\item Window-Abelian-Matching (WM), the prefix based algorithm using the original sliding window approach \cite{GL89,JTU96,Nav97,Ejaz10} with an $O(n)$ worst case complexity;
\item Grabowsky-Faro-Giaquinta (GFG), a prefix based algorithm \cite{GFG11} which uses less branch conditions, with an $O(n)$ worst case complexity;
\item Exact Forward form Small alphabets (EFS), a prefix based adaptation \cite{CGT15} of the BAM algorithm, working in $O(n)$ worst case time complexity for short patterns; 
\item Heap-Counting-Abelian-Matching (HCAM), the prefix based algorithm using the heap counting approach described in Section \ref{new:prefix} and working in $O(n)$ worst case time complexity;
\item Heap-Filtering-Abelian-Matching (HFAM), the filtering variant of the HCAM algorithm, described in Section \ref{new:filter} and working in $O(nm)$ worst case time complexity;
\end{enumerate}
the following 5 suffix based algorithms:
\begin{enumerate}
\item[(6)] Backward-Window-Abelian-Matching (BWM), the suffix based algorithm using the original sliding window approach \cite{Ejaz10} and working inn $O(nm\sigma)$ worst case time complexity;
\item[(7)]  Bit-parallel Abelian Matching algorithm (BAM), a suffix based bit parallel algorithm working for short patterns and implemented using $2$-grams \cite{CF14,CGT15}, with a $O(nm)$ worst case time complexity (in the case of long patterns we used the BAMs implementation \cite{CGT15}); 
\item[(8)]  Exact Backward for Large alphabets (EBL), a simple suffix based bit-parallel algorithm \cite{CGT15} with a $O(nm)$ worst case time complexity;
\item[(9)]  Backward-Heap-Counting-Abelian-Matching (BHCAM), the suffix based algorithm using the heap counting approach described in Section \ref{new:suffix} and working in $O(n)$ worst case time complexity;
\item[(10)]  Backward-Heap-Filtering-Abelian-Matching (BHFAM), the filtering variant of the BHCAM algorithm, described in Section \ref{new:filter} and working in $O(nm)$ worst case time complexity;
\end{enumerate}
and the following SIMD based algorithm:
\begin{enumerate}
\item[(11)]  Equal Any (EA), an efficient prefix based solution \cite{Ghu16} implemented using SIMD instructions;
\end{enumerate}

\smallskip

All algorithms have been implemented in \textsf{C}, and have been tested using the \textsc{Smart} tool~\cite{FTBDM16} and executed locally on a MacBook Pro with 4 Cores, a 2 GHz Intel Core i7 processor, 16 GB RAM 1600 MHz DDR3, 256 KB of L2 Cache and 6 MB of Cache L3.\footnote{The \textsc{Smart} tool is available online at  \url{https://smart-tool.github.io/smart/}.}
Comparisons have been performed in terms of running times, including any preprocessing time.
Experimental evaluations relative to one random sequence and three real data sets are reported in Tables~\ref{exp_tab_general_binary}, \ref{exp_tab_general_genome}, \ref{exp_tab_general_protein}, and \ref{exp_tab_general_english}. 

For our tests on real data, we used a genome sequence, a protein sequence, and an English text (all of size 10MB). In addition we also tested our algorithms on a binary random sequence with a uniform distribution of characters.  Such sequences are provided by the research tool \textsc{Smart}, available online for download (for additional details about the sequences, see the details of the \textsc{Smart} tool \cite{FTBDM16}).  

In the experimental evaluation, patterns of length $m$ were randomly extracted from the sequences, with $m$ ranging over the set of values $\{2^i \mid 1\leq i \leq 8\}$. Thus at least one occurrence is reported for each algorithm execution.
In all cases, the mean over the running times (expressed in hundredths of seconds) of $1000$ runs has been reported.

Tables  \ref{exp_tab_general_binary}, \ref{exp_tab_general_genome}, \ref{exp_tab_general_protein}, and \ref{exp_tab_general_english} summarise the running times of our evaluations. Each table is divided into four blocks. The first block presents the results relative to prefix based solutions, the second block presents the results for the suffix based algorithms, while the third block presents the results for the algorithm based on SIMD instructions. The newly presented algorithms have been marked with a star ($\star$) symbol.

Best results among the two sets of algorithms have been bold-faced to ease their localization, while the overall best results have been also underlined. In addition we included in the last block the speedup (in percentage) obtained by our best newly presented algorithm against the best running time obtained by previous algorithms: positive percentages denote running times worsening, whereas negative values denote performance improvements. Percentages representing performance improvements have been bold-faced.

\newcommand{\newalgo}{\ \ \hspace*{\fill}$\star$}

\begin{table}[!t]
\begin{center}
\textsc{Random binary sequence}
\begin{scriptsize}
\rotatebox[origin=c]{90}{\textsc{\ \ \tiny{simd}\ \ \ \ \ \ \ \ suffix based \ \ \ \ \ \ \ \ \ \ prefix based}}~\begingroup\setlength{\fboxsep}{0pt}
\begin{tabular*}{0.99\textwidth}{@{\extracolsep{\fill}}|l|llllllll|}
\hline
&&&&&&&&\\[-0.2cm]
$m$ & $2$ & $4$ & $8$ & $16$ & $32$ & $64$ & $128$ & $256$\\
&&&&&&&&\\[-0.2cm]
\hline
&&&&&&&&\\[-0.1cm]
\textsc{WM} & 16.66 & 16.75 & 16.36 & 16.31 & 16.31 & 16.46 & 16.44 & 16.29\\
\textsc{GFG} & 23.59 & 23.24 & 23.41 & 23.10 & 23.16 & 23.25 & 23.19 & 23.27\\
\textsc{EFS} & 9.13 & 8.97 & 9.10 & 9.09 & 9.19 & 9.01 & 8.97 & 9.06\\
\textsc{HCAM}\newalgo & \best{7.62} & \best{7.66} & \best{7.62} & \best{7.50} & \best{7.47} & \best{7.70} & \best{7.35} & \best{7.57}\\
\textsc{HFAM}\newalgo & 49.40 & 41.04 & 38.77 & 31.93 & 34.69 & 38.80 & 48.63 & 60.23\\
&&&&&&&&\\[-0.1cm]
\hline
&&&&&&&&\\[-0.1cm]
\textsc{BWM} & 88.23 & 80.62 & 84.04 & 78.77 & 96.49 & 115.69 & 165.28 & 233.74\\
\textsc{BAM} & \good{11.29} & 24.60 & 30.00 & 29.99 & 37.12 & 43.26 & 58.52 & 74.58\\
\textsc{EBL} & 70.69 & 112.78 & 165.69 & 231.29 & - & - & - & -\\
\textsc{BHCAM}\newalgo & 28.82 & \good{22.12} & \good{18.66} & \good{14.54} & \good{13.13} & \good{11.46} & \good{10.78} & \good{10.12}\\
\textsc{BHFAM}\newalgo & 57.60 & 43.43 & 40.02 & 33.62 & 36.60 & 40.34 & 51.26 & 63.62\\
&&&&&&&&\\[-0.1cm]
\hline
&&&&&&&&\\[-0.1cm]
\textsc{EA} & 10.23 & 11.07 & 11.29 & 11.32 & - & - & - & -\\
&&&&&&&&\\[-0.1cm]
\hline
&&&&&&&&\\[-0.1cm]
\emph{Speed-Up} & \good{-13.79\%} & \good{-11.75\%} & \good{-13.47\%} & \good{-14.45\%} & \good{-15.81\%} & \good{-11.80\%} & \good{-14.53\%} & \good{-13.50\%}\\
&&&&&&&&\\
\hline
\end{tabular*}\endgroup
\end{scriptsize}
\end{center}
\caption{\label{exp_tab_general_binary}Experimental results on a random binary sequence.}
\end{table}

\begin{table}[!t]
\begin{center}
\textsc{Genome sequence}
\begin{scriptsize}
\rotatebox[origin=c]{90}{\textsc{\ \ \tiny{simd}\ \ \ \ \ \ \ \ suffix based \ \ \ \ \ \ \ \ \ \ prefix based}}~\begingroup\setlength{\fboxsep}{0pt}
\begin{tabular*}{0.99\textwidth}{@{\extracolsep{\fill}}|l|llllllll|}
\hline
&&&&&&&&\\[-0.2cm]
$m$ & $2$ & $4$ & $8$ & $16$ & $32$ & $64$ & $128$ & $256$\\
&&&&&&&&\\[-0.2cm]
\hline
&&&&&&&&\\[-0.1cm]
\textsc{WM} & 13.63 & 13.02 & 13.09 & 13.04 & 13.04 & 13.06 & 12.97 & 12.88\\
\textsc{GFG} & 20.47 & 20.19 & 20.42 & 20.41 & 20.47 & 20.30 & 20.39 & 20.37\\
\textsc{EFS} & 8.26 & 8.31 & 8.35 & 8.34 & 8.36 & - & - & -\\
\textsc{HCAM}\newalgo& \good{6.97} & \good{6.86} & \good{6.92} & \good{6.93} & \best{6.91} & \best{6.89} & \best{6.89} & \best{6.85}\\
\textsc{HFAM}\newalgo& 20.14 & 11.86 & 9.18 & 7.87 & 7.47 & 7.09 & 7.29 & 7.42\\
&&&&&&&&\\[-0.1cm]
\hline
&&&&&&&&\\[-0.1cm]
\textsc{BWM} & 65.59 & 46.81 & 33.47 & 25.79 & 22.94 & 20.67 & 20.81 & 24.95\\
\textsc{BAM} & \good{10.62} & \good{10.96} & 13.20 & 11.87 & 10.69 & 9.85 & 9.33 & 10.25\\
\textsc{EBL} & 29.95 & 27.44 & 55.69 & 119.26 & 227.14 & - & - & -\\
\textsc{BHCAM}\newalgo & 26.82 & 18.03 & \good{9.98} & \good{7.74} & \good{7.55} & \good{7.32} & \good{7.42} & \good{7.36}\\
\textsc{BHFAM}\newalgo & 37.21 & 21.72 & 11.50 & 9.35 & 9.09 & 8.88 & 8.81 & 9.00\\
&&&&&&&&\\[-0.1cm]
\hline
&&&&&&&&\\[-0.1cm]
\textsc{EA} & \best{4.01} & \best{4.03} & \best{4.56} & \best{4.67} & - & - & - & -\\
&&&&&&&&\\[-0.1cm]
\hline
&&&&&&&&\\[-0.1cm]
\emph{Speed-Up} & +11.87\% & +11.40\% & +10.76\% & +10.55\% & \good{-12.12\%} & \good{-29.16\%} & \good{-22.77\%} & \good{-34.85\%}\\
&&&&&&&&\\
\hline
\end{tabular*}\endgroup
\end{scriptsize}
\end{center}
\caption{\label{exp_tab_general_genome}Experimental results on a genome sequence.}
\end{table}

Consider first the case of small alphabets, and specifically abelian string matching on strings over an alphabet of size $\sigma\leq 4$ (Tables \ref{exp_tab_general_binary} and \ref{exp_tab_general_genome}). 

From experimental results it turns out that, in the case of small alphabets, prefix based solutions are more flexible and efficient than suffix based algorithms. This is because the shift advancements performed by suffix based solutions do not compensate the number of character inspections performed during each iteration. Thus, while prefix based algorithms  maintain a linear behaviour which do not depend on the pattern length, suffix based solutions shown an increasing trend (or a slightly decreasing trend), while the length of the pattern increases, but with a very low performances on average. 
The only exception is the EA algorithm which takes advantage of the efficiency of SIMD instructions outperforming all other solutions in the case of very short patterns and $\sigma\leq 4$. Unfortunately a disadvantage of the EA algorithm is that it works only for $m\leq 16$.

In general the best solutions are obtained by the HCAM algorithm and, as expected, the EFS algorithm, which is specifically tuned to run faster on small alphabets.

\begin{table}[!t]
\begin{center}
\textsc{Protein sequence}
\begin{scriptsize}
\rotatebox[origin=c]{90}{\textsc{\ \ \tiny{simd}\ \ \ \ \ \ \ \ suffix based \ \ \ \ \ \ \ \ \ \ prefix based}}~\begingroup\setlength{\fboxsep}{0pt}
\begin{tabular*}{0.99\textwidth}{@{\extracolsep{\fill}}|l|llllllll|}
\hline
&&&&&&&&\\[-0.2cm]
$m$ & $2$ & $4$ & $8$ & $16$ & $32$ & $64$ & $128$ & $256$\\
&&&&&&&&\\[-0.2cm]
\hline
&&&&&&&&\\[-0.1cm]
\textsc{WM} & 19.64 & 19.43 & 18.94 & 19.22 & 19.04 & 19.00 & 19.14 & 19.20\\
\textsc{GFG} & 30.20 & 29.83 & 30.26 & 30.20 & 30.03 & 29.80 & 29.44 & 30.03\\
\textsc{EFS} & 12.52 & 12.56 & - & - & - & - & - & -\\
\textsc{HCAM}\newalgo & \good{11.04} & \good{10.86} & \good{10.80} & \good{10.81} & \good{10.83} & 10.80 & 10.78 & 10.82\\
\textsc{HFAM}\newalgo & 16.23 & 14.44 & 12.95 & 12.18 & 11.05 & \good{10.49} & \good{10.30} & \good{10.11}\\
&&&&&&&&\\[-0.1cm]
\hline
&&&&&&&&\\[-0.1cm]
\textsc{BWM} & 62.77 & 37.32 & 24.04 & 16.65 & 11.21 & 8.09 & \best{7.09} & \best{6.96}\\
\textsc{BAM} & 16.27 & 13.77 & \best{5.74} & \best{4.74} & - & - & - & -\\
\textsc{EBL} & 16.83 & \good{9.44} & 8.78 & 8.76 & 15.57 & - & - & -\\
\textsc{BHCAM}\newalgo & \good{14.94} & 12.00 & 11.22 & 9.84 & \best{7.90} & \best{7.72} & 9.15 & 9.71\\
\textsc{BHFAM}\newalgo & 16.17 & 12.38 & 11.66 & 9.89 & 7.98 & 7.77 & 9.64 & 10.12\\
&&&&&&&&\\[-0.1cm]
\hline
&&&&&&&&\\[-0.1cm]
\textsc{EA} & \best{6.78} & \best{6.34} & 6.87 & 6.60 & - & - & - & -\\
&&&&&&&&\\[-0.1cm]
\hline
&&&&&&&&\\[-0.1cm]
\emph{Speed-Up} & +28.88\% & +28.66\% & +29.04\% & +24.17\% & \good{-37.11\%} & \good{-2.99\%} & +14.61\% & +19.14\%
\\
&&&&&&&&\\
\hline
\end{tabular*}\endgroup
\end{scriptsize}
\end{center}
\caption{\label{exp_tab_general_protein}Experimental results on a protein sequence.}
\end{table}

\begin{table}[!t]
\begin{center}
\textsc{Natural language text}
\begin{scriptsize}
\rotatebox[origin=c]{90}{\textsc{\ \ \tiny{simd}\ \ \ \ \ \ \ \ suffix based \ \ \ \ \ \ \ \ \ \ prefix based}}~\begingroup\setlength{\fboxsep}{0pt}
\begin{tabular*}{0.99\textwidth}{@{\extracolsep{\fill}}|l|llllllll|}
\hline
&&&&&&&&\\[-0.2cm]
$m$ & $2$ & $4$ & $8$ & $16$ & $32$ & $64$ & $128$ & $256$\\
&&&&&&&&\\[-0.2cm]
\hline
&&&&&&&&\\[-0.1cm]
\textsc{WM} & 17.45 & 17.61 & 17.60 & 17.71 & 17.34 & 17.65 & 17.51 & 17.65\\
\textsc{GFG} & 27.13 & 27.02 & 27.18 & 27.31 & 27.26 & 27.23 & 27.21 & 27.51\\
\textsc{EFS} & 11.88 & 11.39 & - & - & - & - & - & -\\
\textsc{HCAM}\newalgo & \good{10.14} & \good{10.01} & 10.18 & 10.03 & 10.12 & 10.10 & 9.98 & 10.11\\
\textsc{HFAM}\newalgo & 12.20 & 10.11 & \good{9.33} & \good{9.06} & \good{8.84} & \good{8.70} & \good{8.75} & \good{8.57}\\
&&&&&&&&\\[-0.1cm]
\hline
&&&&&&&&\\[-0.1cm]
\textsc{BWM} & 60.05 & 35.18 & 22.53 & 15.17 & 10.09 & 6.82 & 5.30 & 4.92\\
\textsc{BAM2} & 14.45 & 12.01 & \best{5.44} & - & - & - & - & -\\
\textsc{EBL} & 15.30 & \good{9.54} & 7.53 & \good{7.06} & \best{6.13} & 6.11 & 6.51 & 12.38\\
\textsc{BHCAM}\newalgo & \good{14.40} & 12.20 & 10.29 & 8.53 & 6.38 & \best{5.18} & 4.46 & 4.44\\
\textsc{HFAM}\newalgo & 16.28 & 12.63 & 10.66 & 8.59 & 6.40 & 5.23 & \best{4.42} & \best{4.28}\\
&&&&&&&&\\[-0.1cm]
\hline
&&&&&&&&\\[-0.1cm]
\textsc{EA} & \best{6.34} & \best{6.12} & 6.20 & \best{6.98} & - & - & - & -\\
&&&&&&&&\\[-0.1cm]
\hline
&&&&&&&&\\[-0.1cm]
\emph{Speed-Up} & +24.09\% & +23.81\% & +25.79\% & +10.38\% & +1.53\% & \good{-5.68\%} & \good{-4.66\%} & \good{-3.15\%}
\\
&&&&&&&&\\
\hline
\end{tabular*}\endgroup
\end{scriptsize}
\end{center}
\caption{\label{exp_tab_general_english}Experimental results on a natural language text.}
\end{table}

Consider now the case of large alphabets, and specifically abelian string matching on strings over alphabets of size $\sigma>20$ (Tables \ref{exp_tab_general_protein} and \ref{exp_tab_general_english}). 

In contrast with previous results, from our tests it turns out that, in this case, suffix based solutions are more efficient than prefix based algorithms. This is because the shift advancements performed by suffix based solutions substantially exceed the number of character inspections performed during each iteration, allowing a sub.linear behaviour in practice. 
In any case prefix based algorithms show a linear behaviour which do not depend on the pattern length and are more efficient for very short patterns, where the shift proposed by any suffix based solution is too small.
Also in the case of large alphabets, when $m\leq 16$, the EA algorithm outperforms all other algorithms taking advantage of the efficiency of SIMD instructions. Among the faster solutions we mention the BHFAM algorithm, for the case of long patterns, and the EBL algorithm which is specifically tuned for this setting.

\section{Acknowledgement}
We would like to thank Sukhpal Singh Ghuman and Jorma Tarhio for sharing their ideas and the source codes of their algorithms.
We would like to thank also Domenico Cantone for his help and advice.

\bibliographystyle{plain}

\end{document}